\documentclass[journal=nalefd,manuscript=letter]{achemso}

\usepackage[version=3]{mhchem} 
\usepackage[T1]{fontenc}       
\usepackage[]{graphicx}
\usepackage[utf8]{inputenc}
\usepackage[T1]{fontenc}%
\usepackage{lmodern}%
\usepackage{amsmath, amssymb}
\usepackage{hyperref}%
\usepackage{xcolor}%
\usepackage{notes2bib}

\newcommand{\beq}{\begin{equation}\begin{aligned}}
\newcommand{\eeq}{\end{aligned}\end{equation}}

\newcommand{\mum}{$\mu$m}
\newcommand{\ws}{WS\ensuremath{_2}}

\newcommand{\Kplus}{\textsf{\textbf K}}
\newcommand{\Kminus}{\textsf{-\textbf K}}

\newcommand{\figref}[2]{\ref{#1}\textsf{\bfseries #2}}

\definecolor{linkcol}{rgb}{0,0,0.4}
\definecolor{citecol}{rgb}{0.5,0,0}

\hypersetup{
	bookmarksopen=true,
	pdftitle="Microscopic origin of the valley Hall effect in transition metal dichalcogenides revealed by wavelength dependent mapping",
	pdfauthor="N. Ubrig et al.",
	pdfsubject="Microscopic origin of the valley Hall effect in transition metal dichalcogenides revealed by wavelength dependent mapping", 
	pdfstartview={FitH},    
	pdfmenubar=true, 
	pdfhighlight=/O, 
	colorlinks=true, 
	pdfpagemode=UseNone, 
	pdfpagelayout=SinglePage, 
	pdffitwindow=true, 
	linkcolor=linkcol, 
	citecolor=linkcol, 
	urlcolor=blue
}

	\author{Nicolas Ubrig}
	\affiliation{DQMP, Université de Genève, 24 quai Ernest Ansermet, CH-1211, Geneva, Switzerland}
	\alsoaffiliation{GAP, Université de Genève, 24 quai Ernest Ansermet, CH-1211, Geneva, Switzerland}
	\altaffiliation{Contributed equally to this work}
	\email{nicolas.ubrig@unige.ch}
	\author{Sanghyun Jo}
	\affiliation{DQMP, Université de Genève, 24 quai Ernest Ansermet, CH-1211, Geneva, Switzerland}
	\alsoaffiliation{GAP, Université de Genève, 24 quai Ernest Ansermet, CH-1211, Geneva, Switzerland}
	\altaffiliation{Contributed equally to this work}
	\author{Marc Philippi}
	\affiliation{DQMP, Université de Genève, 24 quai Ernest Ansermet, CH-1211, Geneva, Switzerland}
	\alsoaffiliation{GAP, Université de Genève, 24 quai Ernest Ansermet, CH-1211, Geneva, Switzerland}
	\author{Davide Costanzo}
	\affiliation{DQMP, Université de Genève, 24 quai Ernest Ansermet, CH-1211, Geneva, Switzerland}
	\alsoaffiliation{GAP, Université de Genève, 24 quai Ernest Ansermet, CH-1211, Geneva, Switzerland}
	\author{Helmuth Berger}
	\affiliation{Institut de Physique de la Matière Condensée, Ecole Polytechnique Fédérale de Lausanne, CH-1015, Lausanne, Switzerland}
	\author{Alexey B. Kuzmenko}
	\affiliation{DQMP, Université de Genève, 24 quai Ernest Ansermet, CH-1211, Geneva, Switzerland}
	\author{Alberto F. Morpurgo}
	\email{Alberto.morpurgo@unige.ch}
	\affiliation{DQMP, Université de Genève, 24 quai Ernest Ansermet, CH-1211, Geneva, Switzerland}
	\alsoaffiliation{GAP, Université de Genève, 24 quai Ernest Ansermet, CH-1211, Geneva, Switzerland}
	\title{\texorpdfstring{Microscopic Origin of the Valley Hall Effect in Transition Metal Dichalcogenides Revealed by Wavelength Dependent Mapping}{Microscopic origin of the valley Hall effect in transition metal dichalcogenides revealed by wavelength dependent mapping}} 
	
\begin{document}

%
%
%
	
\begin{abstract}
The band structure of many semiconducting monolayer transition metal dichalcogenides (TMDs) possesses two degenerate valleys, with equal and opposite Berry curvature. It has been predicted that, when illuminated with circularly polarized light, interband transitions generate an unbalanced non-equilibrium population of electrons and holes in these valleys, resulting in a finite Hall voltage at zero magnetic field when a current flows through the system. This is the so-called valley Hall effect that has recently been observed experimentally. Here, we show that this effect is mediated by photo-generated neutral excitons and charged trions, and not by inter-band transitions generating independent electrons and holes. We further demonstrate an experimental strategy, based on wavelength dependent spatial mapping of the Hall voltage, which allows the exciton and trion contributions to the valley Hall effect to be discriminated in the measurement. These results represent a significant step forward in our understanding of the microscopic origin of photo-induced valley Hall effect in semiconducting transition metal dichalcogenides, and demonstrate experimentally that composite quasi-particles, such as trions, can also possess a finite Berry curvature.
\end{abstract}

In the presence of finite Berry curvature, charge carriers in an electronic band gain velocity perpendicular to an accelerating electric field, similarly to what happens when an external magnetic field is applied\cite{xiao_valley-contrasting_2007,nagaosa_anomalous_2010,xiao_berry_2010}. As a result, a transverse Hall voltage can appear in the presence of current flow even at zero magnetic field ( $B$=0 T)\cite{xiao_valley-contrasting_2007,yao_valley-dependent_2008}. The phenomenon should be visible in semiconducting transition metal dichalcogenides (TMDs) with broken inversion symmetry, such as monolayers of MoS$_2$, MoSe$_2$, WS$_2$ and WSe$_2$. In these systems, however, two valleys (\Kplus{} and \Kminus{}) with equal and opposite Berry curvature are present, whose contributions to the Hall effect exactly cancel out under equilibrium conditions\cite{xiao_coupled_2012,zeng_optical_2013,xu_spin_2014}. It has been predicted that a Hall voltage transverse to the current flow can be observed if a net valley polarization is generated, i.e., if the electronic states in the two valleys are populated differently. This is a manifestation of the so-called valley Hall effect (VHE)\cite{xiao_coupled_2012}.\\

Early theoretical work insightfully discussed how orbital selection rules for optical inter-band transitions allow a valley polarization to be created by illuminating a monolayer TMD with circularly polarized light\cite{xiao_berry_2010}. This idea has been implemented in recent experiments that have led to the first observation of a $B$=0 T Hall voltage in monolayer MoS$_2$ \cite{mak_valley_2014}. The observations have indeed been interpreted in terms of a valley-unbalanced population of independent electrons and holes created by interband transitions. Such an interpretation is however questionable because the great majority of optoelectronic phenomena observed in TMDs --including the reported VHE-- become visible upon illumination with light at the wavelength corresponding to the energy of  excitons (X$_0$) or trions(X$_{\pm}$). This energy is much smaller, by 0.5 eV or more\cite{zhu_giant_2011,ye_probing_2014,chernikov_exciton_2014,kozawa_photocarrier_2014,chernikov_population_2015}, than the energy needed to cause interband transitions so that, under the conditions of the experiment, illumination cannot directly create independent electrons and holes. It follows that the microscopic process responsible for the occurrence of the observed VHE in monolayer TMDs remains to be understood.\\

This letter is devoted to addressing this question using devices based on monolayer WS$_2$ and bilayer 3R-MoS$_2$ crystals lacking inversion symmetry. We start by discussing two microscopic scenarios that explain how the photo-generation of either charged trions or neutral excitons can lead to the occurrence of VHE. The two scenarios can occur at the same time, and we show that they can be detected and discriminated by mapping the VHE signal as a function of position of the laser spot used to photo-excite the system. We also show that when the wavelength corresponding to exciton and trion energy can be identified experimentally --for instance from a splitting in the photoluminescence (PL) spectra-- mapping the Hall voltage at the corresponding wavelength allows the two contributions to the VHE to be identified in a same device.\\

The simplest way to explain the photo-induced VHE at photon energies far below the threshold for interband transitions is by invoking trions, a scenario that has been recently considered theoretically \cite{yu_dirac_2014}. Trions are charged excitons that bind an additional electron or hole and that inherit a Berry curvature from the electron/hole band states out of which they are formed\cite{yu_dirac_2014}. Experimentally, it has been demonstrated that the trion photoluminescence is circular dichroic \cite{mak_tightly_2013,jones_optical_2013}, implying that the same optical selection rules governing inter-band transitions in TMD monolayers also apply for these quasiparticles\cite{xiao_coupled_2012,mak_control_2012,zeng_valley_2012,cao_valley-selective_2012}. It follows that, upon stationary illumination with circularly polarized light, trions can develop a finite valley polarization and contribute with an off-diagonal term to the conductivity tensor. Such a term would result in a finite Hall voltage at $B$=0 T when a current is forced through the system.\\

By the same argument, it may be expected that valley-polarized excitons should not lead to a measurable VHE: even if excitons possess a Berry curvature --as predicted theoretically\cite{yu_anomalous_2015}-- they cannot influence the conductivity tensor directly, because they are neutral. In particular, their charge neutrality implies that exciton accumulation to the side of a TMD monolayer does not cause a transverse electric field and therefore cannot generate any Hall voltage. Nevertheless, photo-generation of excitons may still lead to the occurrence of VHE, albeit only indirectly. Indeed, in the presence of a strong electric field,  excitons may split and release free electrons and holes that, being charged, can result in a measurable VHE if their valley polarization is preserved during the splitting process. An electric field of the required strength can be generated by specific charged defects or near the interface with a metal, due to the formation of the Schottky barrier. We anticipate that in our experiments we will rely on the possibility to split exciton at interfaces between TMDs and metal contacts, which has been already demonstrated experimentally in TMDs (see, for instance, Refs. \citenum{wu_elucidating_2013,ubrig_scanning_2014}).\\

\begin{figure*}[t]
	\centering
	\includegraphics[width=0.8\textwidth]{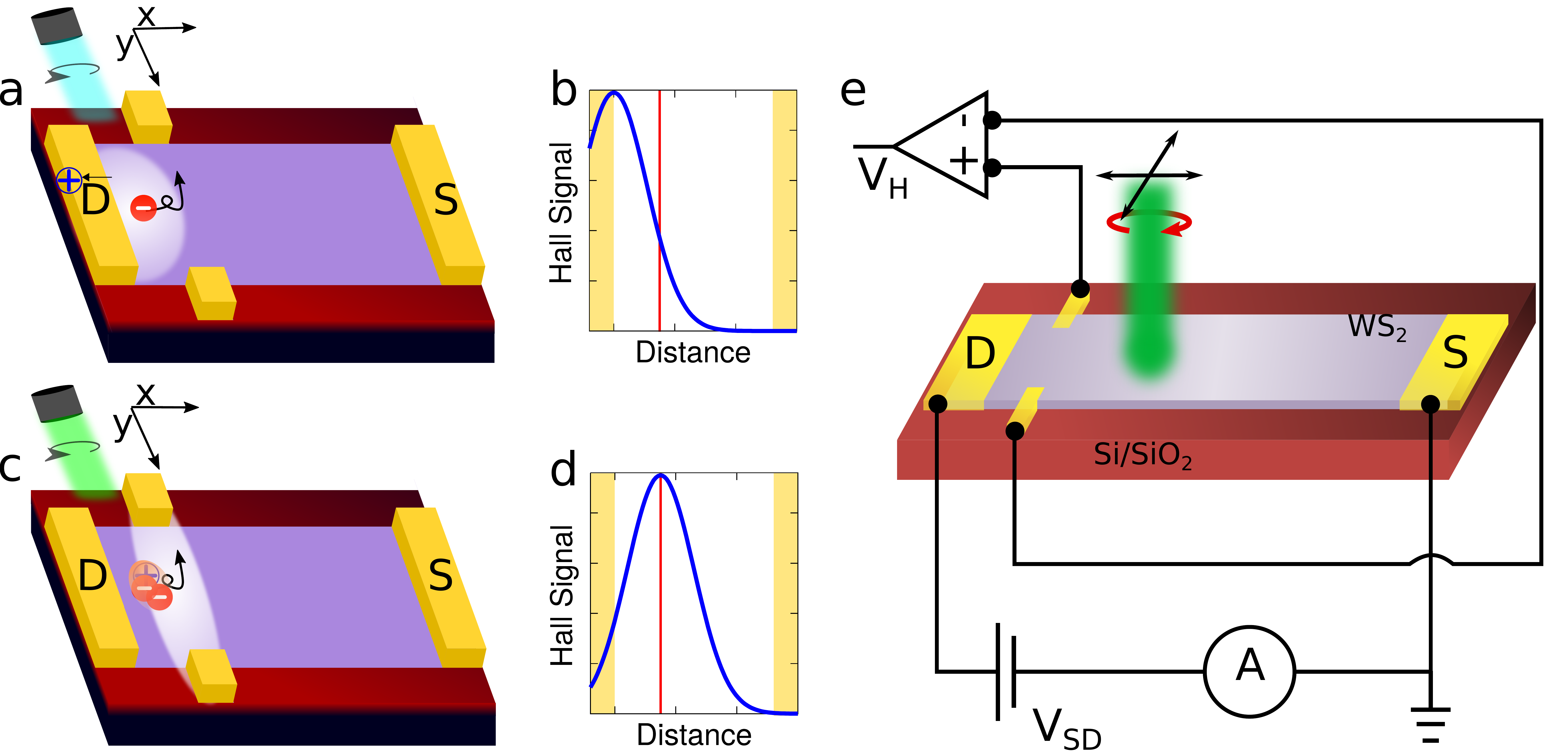}%
	\caption{Wavelength dependent mapping of the VHE  in a long monolayer TMD transistor. If the Hall probes are positioned close to one of the contacts, the contribution of either trions or excitons can be detected by measuring the Hall voltage upon scanning the laser spot. Panels \textsf{\bfseries a,b} illustrate what is expected when the laser is tuned at a wavelength corresponding to the photogeneration of neutral excitons. Excitons need to split at the D contact (\textsf{\bfseries a}) to generate charged carriers that can modify the conductivity tensor and contribute to the VHE. Hence, the Hall voltage peaks when the laser spot is focused at the interface between the TMD monolayer and the drain contact (as shown in \textsf{\bfseries b}). Panels \textsf{\bfseries c,d} represent the situation when the laser light is tuned at a wavelength corresponding to the photogeneration of trions. Charged trions  (\textsf{\bfseries c}) directly create an off-diagonal contribution to the conductivity which leads to a maximum signal when the laser spot is focused in the middle of the Hall probes, as indicated by the profile shown in \textsf{\bfseries d} (in panels \textsf{\bfseries b} and \textsf{\bfseries d} the yellow regions represent the drain (D) and source (S) contacts; the red line marks the center of the Hall probes). \textsf{\bfseries e}	Schematic representation of the electrical circuit used to bias the device and measure the Hall voltage as the laser spot is scanned around. }
	\label{fig:mapsketch}
\end{figure*}

These two distinct microscopic mechanisms responsible for a photo-induced VHE at sub-gap wavelength are not mutually exclusive and can occur at the same time. It is however possible to discriminate between them in a properly designed device configuration by mapping the spatial dependence of the VHE at different laser wavelengths (i.e., by  mapping the Hall voltage as a function of the position of the laser spot used to illuminate the device). With reference to the device geometry shown in Figure \figref{fig:mapsketch}{a}, if the wavelength is tuned to the exciton absorption energy, we expect the Hall signal to be maximum when the laser is focused next to the edge of the current-injecting contact (Figures \figref{fig:mapsketch}{a} and \figref{fig:mapsketch}{b}). That is because, in order to contribute to the Hall voltage,  photo-generated excitons have to reach the interface with the metallic contact, where the electric field is strong enough to split them. The situation is fully analogous to what is observed in scanning photocurrent microscopy measurements\cite{ahn_scanning_2005,wu_elucidating_2013,ubrig_scanning_2014,yamaguchi_spatially_2015}: only excitons that split at the contacts --releasing a majority carrier in the TMD with the minority carrier escaping in the metal electrode-- contribute to the photocurrent. Indeed, in the experiments a photocurrent signal is observed only when the laser spot is focused within the exciton recombination length from the contact, approximately one micron in TMD monolayers\cite{wang_ultrafast_2012,ubrig_scanning_2014,yamaguchi_spatially_2015,moody_exciton_2016}. As compared to photocurrent measurements, the observation of the VHE imposes one more constraint, because building up a Hall voltage requires a finite valley polarization. Therefore, to maximize the measured Hall signal, the transverse electrodes used as Hall probes need to be sufficiently close to the electrode where excitons split, (the characteristic length is the inter-valley scattering length, which is of the order of 1 \mum{}\cite{xu_spin_2014}).\\

It follows from these considerations that --when illumination occurs at the wavelength corresponding to exciton absorption-- the profile of the measured Hall voltage as the laser spot position is scanned through the device is similar to the one sketched in Figure \figref{fig:mapsketch}{b}: the signal peaks when the laser is focused at the interface with the metal contact and decreases as the laser spot is moved into the TMD layer, away from the contact. The profile asymmetry relative to the center of the Hall probes is a direct manifestation of the need to split the photo-generated excitons to create charged carriers. This behavior is different from what is expected when the laser wavelength is tuned to photo-generate trions (see Figure \figref{fig:mapsketch}{c} and  \figref{fig:mapsketch}{d}). In that case the situation is conceptually simpler, because the photo-induced charged trions directly cause a change in the local conductivity tensor, and the contact used to inject current plays no role. Therefore, for the trion contribution to the VHE we expect the profile of the measured Hall voltage to be similar to the one shown in Figure \figref{fig:mapsketch}{d}: the signal is largest when the laser spot is focused in the center of the Hall probes and decays as the laser beam is scanned away on either side.\\

Both contributions to the VHE can be detected in a same device by mapping the profile of the Hall voltage as a function of laser spot position, while selecting the laser wavelength to mainly generate either excitons or trions  (for a schematic representation of experimental configuration employed to perform the wavelength dependent mapping see Figure \figref{fig:mapsketch}{e}). In the devices investigated so far, \cite{mak_valley_2014} this strategy could not be implemented effectively, because the separation between source and drain contacts was too small, resulting in a virtually uniform device illumination. Nevertheless, despite not being ideal for a mapping experiment, a device with dimensions comparable to the intervalley scattering length is advantageous to maximize valley polarization, and hence the Hall voltage. That is why, before mapping the VHE in larger devices, we measured the VHE on a \ws{} monolayer device with small contact separation (for details about the experimental setup, see Supporting Information S1).\\

\begin{figure*}[ht!]
	\centering
	\includegraphics[width=0.8\textwidth]{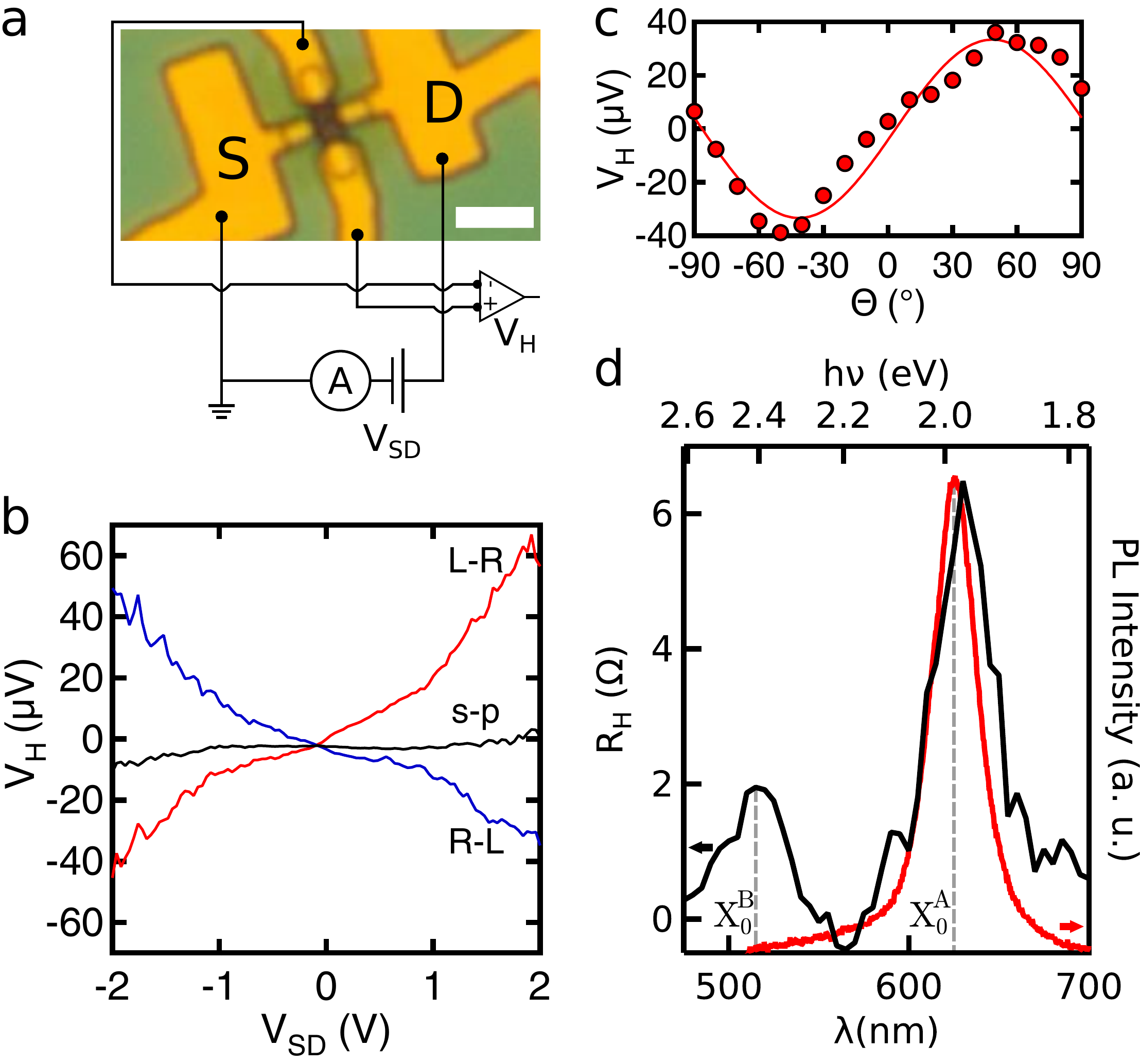}%
	\caption{\textsf{\bfseries a}. Optical micrograph of a short channel monolayer WS$_2$ FET, with the schematics of the electrical circuit used to measure the VHE (the scale bar is 2 \mum{}). \textsf{\bfseries b}. Bias dependence of the Hall voltage, V$_{\textsf H}$, induced in \ws{} upon uniform illumination with circularly polarized light at $\lambda$ = 615 nm. The data is measured with a lock-in amplifier upon periodically switching the polarization state between L and R. The signal changes sign if the polarization sequence is inverted (from L-R to R-L), and disappears for linearly polarized light(s-p), as expected. \textsf{\bfseries c}. Dependence of V$_{\textsf H}$ on the angle $\theta$ between the linear polarization of the light and the optical axis of the photoeleastic modulator used to generate circular polarization (data measured at V$_{\textsf{SD}}$ = +1 V; the red line is a fit with a sinusoidal dependence). \textsf{\bfseries d}. Spectral dependence of the measured Hall resistance R$_{\textsf{H}}$ = V$_{\textsf{SD}}$ /I$_{\textsf{SD}}$ \cite{mak_valley_2014} (black line; data taken at V$_{\textsf{SD}}$ = +1 V). The grey dashed lines mark the position of the A (X$_0^A$) and B (X$_0^B$) exciton, respectively. The red curve is the photoluminescence spectrum measured with a 488 nm laser excitation.}
	\label{fig:spectraldep}
\end{figure*}

An optical image of a \emph{small} device is shown in Figure \figref{fig:spectraldep}{a}, together with a schematic of the electrical circuit used to measure the Hall voltage. It consists in a field-effect transistor (FET) realized using a monolayer WS$_2$ exfoliated onto a heavily doped silicon substrate (acting as gate) covered by thermally grown SiO$_2$. Gold contacts are realized by conventional electron-beam lithography, metal evaporation and lift-off. The separation between the source (S) and drain (D) contacts  is less than 2 \mum{}, just slightly larger than the diameter of the optical beam ($\approx$ 1 \mum{}). Over the years, we have studied the transport properties of FETs realized with mono and multilayers of a variety of different TMDs and demonstrated good control of their transport properties. Using mono, bi, and multi-layers of both WS$_2$ and MoS$_2$, for instance, we have succeeded in observing gate induced superconductivity\cite{jo_electrostatically_2015,costanzo_gate-induced_2016}, ambipolar transport of sufficient quality to determine the band gap\cite{braga_quantitative_2012,ponomarev_ambipolar_2015}, and in realizing light emitting transistors\cite{jo_mono-_2014}. The quality of the devices that we have used for the investigation of the VHE is virtually identical to that of the devices used in these past studies, and we refer to the related publications for all general aspects concerning the device electrical characterization. For the experiments done here, we take advantage of photo-doping --namely the fact that prolonged exposure to visible light creates a stable density of electrons in the layer (see Supporting Information S2)-- and operate the devices without applying any gate voltage ($V_G$=0 V). A key reason is that the VHE mapping measurements discussed below require the device to be illuminated constantly for prolonged periods of time (typically one hour per map, or more). Therefore, to avoid that photo-doping occurring during the measurement itself causes the device properties to drift, it is essential to perform systematic electrical measurements only after having kept the device illuminated for a day or longer, i.e., enough to reach saturation of photo-doping. This typically results in electron densities $n$ of the order of $n= 2.5-5 \cdot 10^{12}$ cm$^{-2}$, which enable transport measurements down to low temperature.\\

The measurement procedures that we follow are  virtually identical to those described in Ref. \citenum{mak_valley_2014} (see Supporting Information S1 for more details). We modulate the polarization state of the laser light shed onto the device with a photoelastic modulator (the modulation frequency is 50 kHz for circular and 100 kHz for linear polarization). The laser wavelength is fixed at the value determined by the maximum of the photoluminescence (PL) intensity ($\lambda$ = 615 nm). The Hall voltage is measured in phase with the modulation of the light polarization using a lock-in technique. When the polarization state of the light is periodically modulated  from left-hand circular (L) to right-hand circular (R) a clear signal is observed. The Hall voltage changes its sign upon reversing the modulation sequence from L-R to R-L (Figure \figref{fig:spectraldep}{b}) and its magnitude varies with the sine of twice the angle of the incident linear polarization with respect to the optical axis of the photoelastic modulator, as expected (Figure \figref{fig:spectraldep}{c}; all data are taken at T = 80 K if not specified otherwise, see section S1 in the Supporting Information). Note that the modulation between two orthogonal linear polarizations (s-p) gives no effect. The data reproduces in WS$_2$ monolayers the observations made earlier by  Mak {\it et al.}\cite{mak_valley_2014} on MoS$_2$. The wavelength at which the phenomenon is observed (Figure \figref{fig:spectraldep}{d}) confirms that, contrary to the interpretation given in Ref. \citenum{mak_valley_2014}, the VHE cannot be mediated by inter-band transitions (Figure \figref{fig:spectraldep}{d})\cite{ye_probing_2014,chernikov_exciton_2014,chernikov_population_2015}.\\

\begin{figure*}[ht!]
	\centering
	\includegraphics[width=0.95\textwidth]{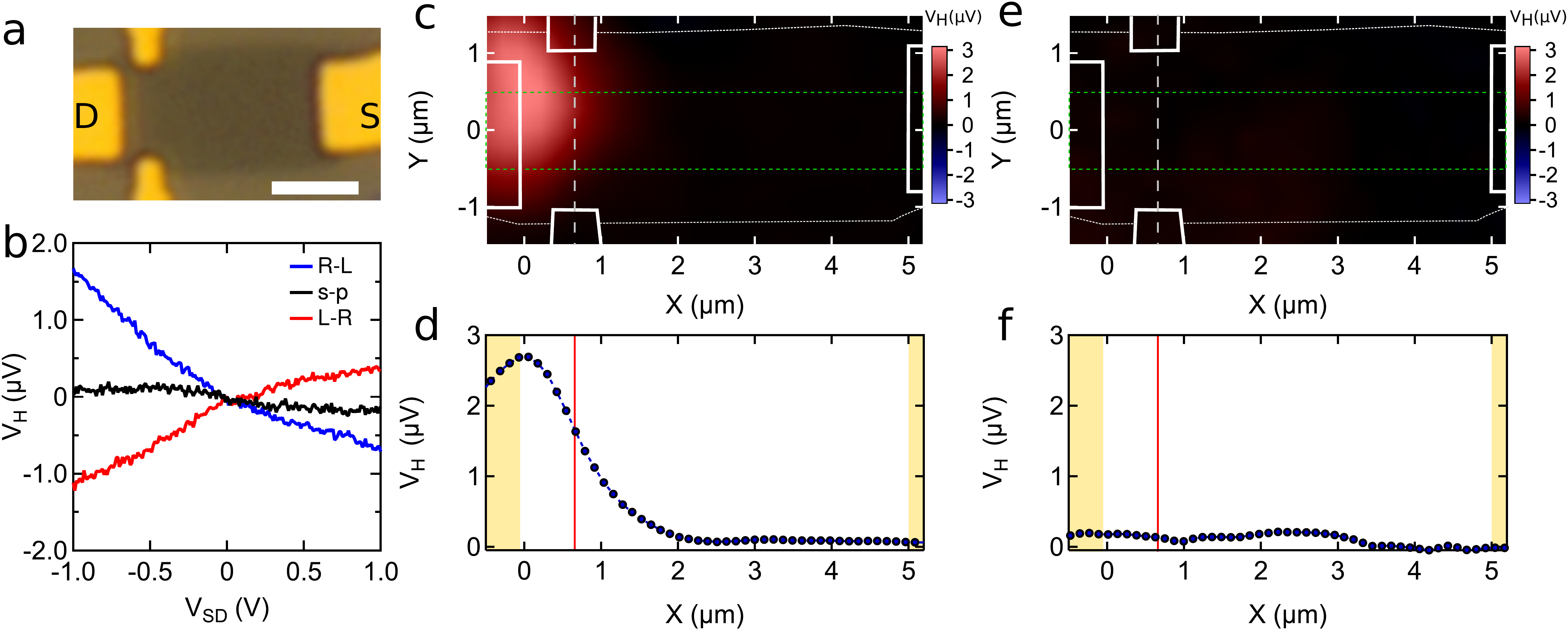}%
	\caption{\textsf{\bfseries a}. Optical micrograph of a \ws{} monolayer device used for wavelength dependent spatial mapping of the Hall voltage. The scale bar is 2 \mum{}. \textsf{\bfseries b}. Bias dependence of the Hall voltage measured when the laser beam is focused close to the drain (D) contact (data taken upon modulating the light polarization state, as discussed in the main text). \textsf{\bfseries c}. Spatial map of the Hall voltage V$_{\rm H}$ measured as a function of laser beam position (data taken with a R-L modulation sequence at V$_{\rm SD}$ = - 1 V and $\lambda$ = 615 nm). The white lines represent the contour of the metallic electrodes and the grey dotted lines the edges of the WS$_2$ monolayer. \textsf{\bfseries d}. Hall voltage as a function of the distance (X) from the drain contact. The profile is obtained by averaging the data in \textsf{\bfseries b}, in the Y interval delimited by green dashed lines (to minimize direct overlap between the laser spot and the Hall probes, as well as  with the sample edges). The Hall signal peaks when the laser is focused near to the D contact, and is asymmetric relative to the center of the Hall probes (marked by the red line). Panels \textsf{\bfseries e} and \textsf{\bfseries f} show the same measurements as in panels \textsf{\bfseries b} and \textsf{\bfseries d} but with  linear, rather than circular, light polarization (s-p modulation sequence; V$_{\rm SD}$ = - 1 V and $\lambda$ = 615 nm). No Hall voltage is seen irrespective of the laser position, as expected.}
	\label{fig:vhews2}
\end{figure*}

Spatially resolved measurements of the VHE (i.e., VHE mapping) were first performed on monolayer WS$_2$  devices with a considerably larger source-drain separation. As discussed above, the Hall probes are close to one of the contacts, within a distance of approximately the inter-valley relaxation length (see Figure \figref{fig:mapsketch}{e} for a schematic representation and Figure \figref{fig:vhews2}{a} for an actual image of one of the devices). The devices are mounted in a cryostat with optical access, under a microscope used to focus the laser, on a piezo-driven stage that enables scanning the device position relative to the fixed laser spot. The same detection technique described above is used in the VHE mapping experiments. Figure \figref{fig:vhews2}{b} shows that upon changing the polarization sequence, the evolution of the bias dependence of the Hall signal exhibits the same behavior as in the uniformly illuminated device (data taken with the laser spot focused at the interface with the drain contact). The spatial map of Hall voltage obtained upon biasing the device with V$_{\rm SD}$ = -1 V is shown in Figure \figref{fig:vhews2}{c} (image taken with  $\lambda$ = 615 nm). Albeit the absolute amplitude of the signal is much smaller than in the shorter channel WS$_2$ device discussed earlier, it is apparent that the signal is more intense when the laser spot is at the interface with the drain contact (as for the absolute magnitude of the signal, we found a strong  dependence on the specific device measured, whose origin is not currently clear). The key aspect of these measurements is summarized in Figure \figref{fig:vhews2}{d} that shows how the spatial profile of the measured Hall voltage decays monotonously as the laser spot is scanned into the WS$_2$ monolayer, away from the interface with the drain  contact. In particular, it is apparent that the spatial dependence is not symmetric relative to the center of the Hall probes, whose position is marked by the red thin line. This is precisely the behavior expected in the case in which the dominant contribution to the VHE is due to valley polarized independent electrons originating from splitting of excitons. Measurements on two additional devices exhibiting the same behavior are shown in the Supporting Information, section S4.\\
As a consistency check, we also repeated the mapping experiment by modulating the incident light between two linearly polarized orthogonal states. Linearly polarized light induces transitions in the \Kplus{} and \Kminus{} valley with equal probability and causes no valley imbalance. Accordingly, this should not lead to the appearance of any Hall voltage. Figure \figref{fig:vhews2}{e} shows that this is indeed the case (the color scale is the same as in Figure \figref{fig:vhews2}{b}), and the corresponding Hall voltage profile plotted in Figure \figref{fig:vhews2}{f} confirms that  $V_H \approx 0$ V throughout the device. Besides establishing that the device behaves as expected, these observations are useful to exclude  possible spurious effects. For instance, it may be argued that the Hall signal detected upon illumination with circularly polarized light does not originate from the VHE, and that it is just the consequence of the rectification at the contacts used as Hall probes\bibnote{In simple terms, Hall probes are metal-semiconductor Schottky contacts behaving as diodes, and they can in principle produce a dc signal by rectifying the oscillating high-frequency currents produced by the incident light. An asymmetry in the contact would then lead to a voltage difference between the Hall probes that could be mistakenly interpreted as a manifestation of the VHE.}. The absence of signal when mapping under illumination with linearly polarized light excludes this possibility, because rectification at Schottky contacts is not sensitive to light polarization (i.e., if the Hall voltage that we observe upon illumination with circularly polarized light was due to rectification, we should see essentially the same signal with linearly polarized light).\\

\begin{figure*}[h!]
	\centering
	\includegraphics[width=0.89\textwidth]{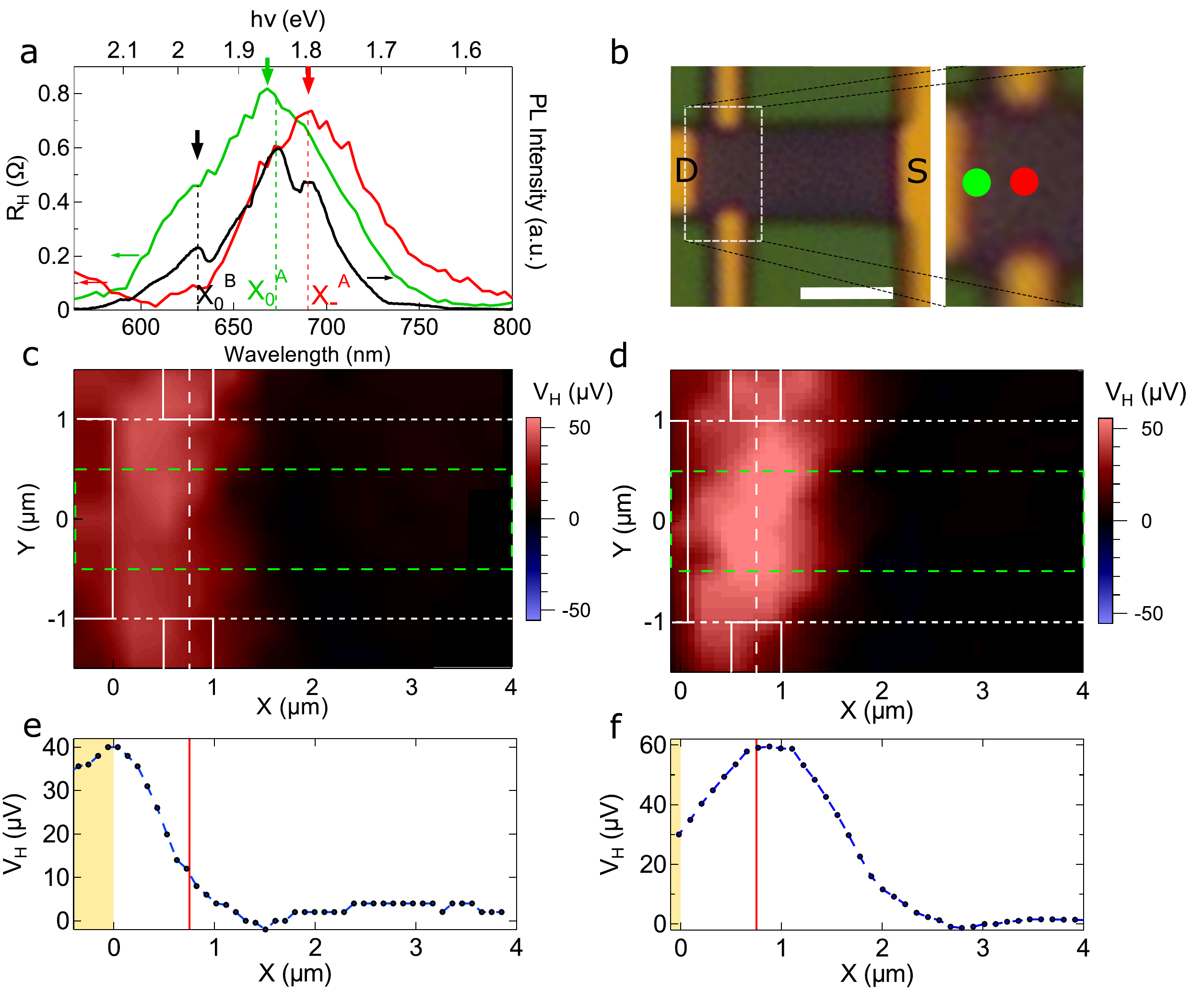}%
	\caption{\textsf{\bfseries a}. Spectral dependence of the Hall resistance R$_{\textsf{H}}$ measured on the bilayer 3R-MoS$_2$ device shown in \textsf{\bfseries b} (the scale bar is 2 \mum{}). The green curve is measured with the laser spot focused near the D contact (green dot in the zoomed out image in \textsf{\bfseries b}) and the red curve with the laser spot focused in the middle of the Hall probes (red dot in the zoomed out image in \textsf{\bfseries b}, V$_{\textrm{SD}}$ = -0.3 V; R-L polarization sequence). The black curve represents the photoluminescence spectrum of the same device. The vertical dashed lines and the arrows indicate the emission wavelengths of the trion (X$_-^A$, red), the A exciton (X$_0^A$, green), and the B exciton (X$_0^B$, black). \textsf{\bfseries c}. Spatial map of the Hall voltage measured with the laser wavelength set on the X$_0^A$ absorption peak ($\lambda$ = 665 nm; V$_{\textrm{SD}}$ = +1 V,  L-R circular polarization sequence). As in Figure \figref{fig:vhews2}{c}, the continuous white lines indicate the edges of the metal contacts, and the dashed lines the edges of the TMD bilayer. \textsf{\bfseries d.} Same as \textsf{\bfseries c}, but with the laser light tuned at the wavelength corresponding to the trion absorption X$_-^A$ ($\lambda$ = 690 nm). Panels \textsf{\bfseries e, f}. Hall signal as a function of the distance (X) from the drain (D) contact for exciton resonance, by averaging the data in the Y interval delimited by the green dashed lines  in \textsf{\bfseries c} and \textsf{\bfseries d}, respectively, corresponding to the area where the laser spot does not directly overlap with the Hall probes and the edges of the device. In \textsf{\bfseries e}, the Hall signal is maximum when the laser spot is focused at the interface with the D contact, as expected for the exciton contribution to the VHE. In \textsf{\bfseries f}, the Hall signal peaks when the laser is focused in the middle of the Hall probes, as expected for the trion contribution.}
	\label{fig:mos2vhe}
\end{figure*}

We now proceed to probing the trion contribution to the VHE, for which we need to tune the laser to match the trion energy. The required wavelength can be determined experimentally if the PL spectrum exhibits a clear splitting between exciton and trion emission. In practice, however, our WS$_2$ devices do not show such a splitting (see Figure \figref{fig:spectraldep}{d}), as it is also the case for many devices reported in the literature\cite{mitioglu_optical_2013,cadiz_excitonic_2017}. Therefore, for this experiment we use a device based on a 3R-stacked  MoS$_2$ bilayer, in which the splitting between excitons and trions could be resolved experimentally (see Figure \figref{fig:mos2vhe}{a}, black curve)\cite{pei_exciton_2015}. Owing to their stacking,  3R TMD multilayers have broken inversion symmetry regardless of the number of layers\cite{suzuki_valley-dependent_2014} and are expected to show a large VHE in all cases. This is a crucial difference from the commonly investigated 2H polytype of TMD multilayers\cite{mak_tightly_2013}, in which the crystal inversion symmetry is strongly broken only in odd multilayers\cite{mak_control_2012,zeng_valley_2012,cao_valley-selective_2012} (in even 2H multilayers, inversion symmetry can still be broken --but much more weakly-- by the presence of an underlying substrate\cite{lee_electrical_2016,wu_electrical_2013}).\\

Prior to recording the  spatially resolved maps, we look at the wavelength dependence of the Hall voltage measured when the laser spot is positioned at the interface with the drain contact, and we compare it to the case in which the laser spot is positioned in the center of the Hall probes (see Figure  \figref{fig:mos2vhe}{b} for the device geometry: the green and red dots in the zoomed out image represent the locations at which the laser spot was centered in the two cases). Notably, the spectral dependence of the Hall voltage in the two cases is  different (see the green and red curves in Figure  \figref{fig:mos2vhe}{a}). When the laser spot is located at the interface with the drain contact, the maximum in Hall voltage occurs at shorter wavelength (green curve in Figure  \figref{fig:mos2vhe}{a}), at a value of $\lambda$ that coincides with exciton peak position in the PL curve. When the laser spot is located in between the Hall probes, the maximum of the Hall voltage occurs at a longer wavelength (see red curve in Figure  \figref{fig:mos2vhe}{a}), very close to the value of  $\lambda$ at which the trion peak is seen in the PL spectrum. This difference in the spectral dependence of the measured Hall voltage represents a first indication that the dominant process responsible for the VHE observed in the two cases is indeed distinct (for additional data see the Supporting Information, section S3).\\
A detailed analysis requires looking at the full maps of the Hall voltage, which we measure first with the laser tuned at a wavelength $\lambda=665$ nm (corresponding to the neutral exciton peak in the PL spectrum), and subsequently at  $\lambda=690$ nm (corresponding to the trion peak in the PL spectrum). The measurements show that the two maps are different (see Figure  \figref{fig:mos2vhe}{c} and  \figref{fig:mos2vhe}{d}). In particular, when the device is photo-excited at the exciton wavelength, the magnitude of the Hall voltage peaks when the laser spot is focused next to the interface between the TMD layer and the metal contact. This behavior is summarized by the profile of the Hall voltage (Figure  \figref{fig:mos2vhe}{e}) extracted from the full map: the maximum $V_H$ value is observed when the laser is positioned at the interface with the metal, and the asymmetry relative to center of the Hall probes is clearly visible, in complete agreement with what we have anticipated for the exciton-mediated contribution to the VHE.  When the Hall voltage map is recorded with the laser tuned at the trion wavelength, on the contrary, the signal peaks when the laser is focused between the Hall probes, as can be clearly seen when looking at the profile plotted in Figure  \figref{fig:mos2vhe}{f}. Note also that the profile is symmetric relative to the center of the Hall probes (as long as the laser is focused onto the TMD layer). The experimental results are consistent with the behaviour anticipated for the case in which the VHE is dominated by the contribution of photo-generated trions (data measured on an additional 3R-MoS$_2$ bilayer device exhibit the same behavior, see section S5 in the Supporting Information). We conclude that mapping the Hall voltage at different wavelength does indeed confirm that photo-generation of both exciton and trion contribute to the occurrence of the VHE, either indirectly or directly, without the need to invoke direct inter-band transitions.\\

With this newly gained insight in the processes responsible for the occurrence of the VHE, we can interpret other details of the experimental results. To this end, we look back at the data shown in figure \figref{fig:spectraldep}{d} measured on a small WS$_2$ device, in which a clear peak is seen in the VHE signal at $\lambda$=520 nm (2.38 eV). In WS$_2$, this energy corresponds to the so-called B-exciton (X$_0^B$) \cite{beal_transmission_1972,beal_kramers-kronig_1976,zhu_giant_2011,zhao_evolution_2013,zeng_optical_2013}, formed by a hole in the lowest energy spin-split valence band and an electron in the conduction band. For a same given valley (experimentally fixed by the circular polarization state of the exciting light), the hole and the electron in a B-exciton have opposite spin as compared to the A-exciton. Since, photo-generated B-excitons also split at nearby metallic contacts and release the majority carrier, B-excitons contribute to the VHE as much as A-excitons do. However, owing to the  opposite spins of the constituents electrons and holes in the B-exciton, the valley-polarized electrons that remain in the TMD layer have opposite spin as compared to electrons generated by splitting A-excitons. Therefore, finding that the measured sign of the Hall voltage is the same irrespective of whether a A- or a B-exciton splits implies that the measured valley Hall effect does not depend on the spin of the charge carriers. That is: what we observe is truly a valley Hall effect and not a spin Hall effect\cite{xiao_coupled_2012}. Similar considerations can be made by looking at the data measured on the 3R-MoS$_2$ bilayers. In the PL of that device (black curve in Figure \figref{fig:mos2vhe}{a}) the B-exciton is visible as a small but distinct peak at $\lambda=630$ nm. The contribution of the B-exciton enhances the magnitude of the VHE, as it is manifested by the shoulder visible at the same wavelength in the spectral dependence of the Hall voltage (green curve in the same figure or Figure S3\textsf{\bfseries c} in the Supporting Information, where the shoulder is more pronounced). Therefore also in the 3R-MoS$_2$ bilayer, as in the WS$_2$ monolayer, the data show that the B-exciton and the A-exciton contribute to the Hall voltage with the same sign (confirming that the phenomenon observed is indeed a valley Hall effect, independent of the spin direction)\cite{xiao_coupled_2012}.\\

There remain important aspects of the valley Hall effect that need to be understood. A relevant question, for instance, is why the magnitude of the effect varies by one order of magnitude in different devices. While possibly an important role is played by differences in the strength and type of disorder, and hence in the value of the inter-valley scattering length (which suppresses the valley polarization and hence the magnitude of the measured Hall voltage) and of the exciton recombination length (which crucially determines the magnitude of the valley Hall effect due to exciton splitting), a better understanding would likely help to reveal important aspects of the microscopic electronic processes that take place in TMDs. Irrespective of these details, we emphasize a conceptual difference between the (indirect) excitonic contribution to the VHE and the direct contributions due to trions,which can be drawn from our work. As we have explained, neutral excitons contribute to the VHE because they mediate the generation of a valley imbalance in the population of majority carriers. This implies that what is actually measured when looking at the excitonic contribution to the VHE is the effect of the Berry curvature of band electrons. For the trion contribution the situation is different, because it is the trion themselves that contribute to the VHE. The detection of a trion contribution in the experiments, therefore, directly implies that trions in TMDs possess a non-trivial Berry curvature\cite{yu_dirac_2014}. Although it had been predicted theoretically, this fact is worth emphasizing explicitly because trions are composite excitations formed by two electrons and a hole (or two holes and one electron), and the experimental detection of the effect of Berry curvature on the dynamics of such a composite quasiparticle had not been reported previously.

\section*{Acknowledgements}
We gratefully acknowledge J. Teyssier, I. Gutierrez, D.-K. Ki, J.-M. Poumirol, D. van der Marel, and A. Ferreira for fruitful discussions and technical help in several occasions. Financial support from the Swiss National Science Foundation, the NCCR QSIT, and the EU Graphene Flagship project is also gratefully acknowledged. NU acknowledges funding from an \emph{Ambizione} grant of the Swiss National Science Foundation.

\section*{Competing financial interests}
The authors declare no competing financial interests.

\begin{suppinfo}
	Experimental procedures and device fabrication; Photo-doping; VHE of Bilayer 3R-MoS$_2$; Additional devices;
\end{suppinfo}

\providecommand{\latin}[1]{#1}
\providecommand*\mcitethebibliography{\thebibliography}
\csname @ifundefined\endcsname{endmcitethebibliography}
{\let\endmcitethebibliography\endthebibliography}{}


\begin{mcitethebibliography}{42}
	\providecommand*\natexlab[1]{#1}
	\providecommand*\mciteSetBstSublistMode[1]{}
	\providecommand*\mciteSetBstMaxWidthForm[2]{}
	\providecommand*\mciteBstWouldAddEndPuncttrue
	{\def\EndOfBibitem{\unskip.}}
	\providecommand*\mciteBstWouldAddEndPunctfalse
	{\let\EndOfBibitem\relax}
	\providecommand*\mciteSetBstMidEndSepPunct[3]{}
	\providecommand*\mciteSetBstSublistLabelBeginEnd[3]{}
	\providecommand*\EndOfBibitem{}
	\mciteSetBstSublistMode{f}
	\mciteSetBstMaxWidthForm{subitem}{(\alph{mcitesubitemcount})}
	\mciteSetBstSublistLabelBeginEnd
	{\mcitemaxwidthsubitemform\space}
	{\relax}
	{\relax}
	
	\bibitem[Xiao \latin{et~al.}(2007)Xiao, Yao, and
	Niu]{xiao_valley-contrasting_2007}
	Xiao,~D.; Yao,~W.; Niu,~Q. \emph{Physical Review Letters} \textbf{2007},
	\emph{99}, 236809\relax
	\mciteBstWouldAddEndPuncttrue
	\mciteSetBstMidEndSepPunct{\mcitedefaultmidpunct}
	{\mcitedefaultendpunct}{\mcitedefaultseppunct}\relax
	\EndOfBibitem
	\bibitem[Nagaosa \latin{et~al.}(2010)Nagaosa, Sinova, Onoda, MacDonald, and
	Ong]{nagaosa_anomalous_2010}
	Nagaosa,~N.; Sinova,~J.; Onoda,~S.; MacDonald,~A.~H.; Ong,~N.~P. \emph{Reviews
		of Modern Physics} \textbf{2010}, \emph{82}, 1539--1592\relax
	\mciteBstWouldAddEndPuncttrue
	\mciteSetBstMidEndSepPunct{\mcitedefaultmidpunct}
	{\mcitedefaultendpunct}{\mcitedefaultseppunct}\relax
	\EndOfBibitem
	\bibitem[Xiao \latin{et~al.}(2010)Xiao, Chang, and Niu]{xiao_berry_2010}
	Xiao,~D.; Chang,~M.-C.; Niu,~Q. \emph{Reviews of Modern Physics} \textbf{2010},
	\emph{82}, 1959--2007\relax
	\mciteBstWouldAddEndPuncttrue
	\mciteSetBstMidEndSepPunct{\mcitedefaultmidpunct}
	{\mcitedefaultendpunct}{\mcitedefaultseppunct}\relax
	\EndOfBibitem
	\bibitem[Yao \latin{et~al.}(2008)Yao, Xiao, and Niu]{yao_valley-dependent_2008}
	Yao,~W.; Xiao,~D.; Niu,~Q. \emph{Physical Review B} \textbf{2008}, \emph{77},
	235406\relax
	\mciteBstWouldAddEndPuncttrue
	\mciteSetBstMidEndSepPunct{\mcitedefaultmidpunct}
	{\mcitedefaultendpunct}{\mcitedefaultseppunct}\relax
	\EndOfBibitem
	\bibitem[Xiao \latin{et~al.}(2012)Xiao, Liu, Feng, Xu, and
	Yao]{xiao_coupled_2012}
	Xiao,~D.; Liu,~G.-B.; Feng,~W.; Xu,~X.; Yao,~W. \emph{Physical Review Letters}
	\textbf{2012}, \emph{108}, 196802\relax
	\mciteBstWouldAddEndPuncttrue
	\mciteSetBstMidEndSepPunct{\mcitedefaultmidpunct}
	{\mcitedefaultendpunct}{\mcitedefaultseppunct}\relax
	\EndOfBibitem
	\bibitem[Zeng \latin{et~al.}(2013)Zeng, Liu, Dai, Yan, Zhu, He, Xie, Xu, Chen,
	Yao, and Cui]{zeng_optical_2013}
	Zeng,~H.; Liu,~G.-B.; Dai,~J.; Yan,~Y.; Zhu,~B.; He,~R.; Xie,~L.; Xu,~S.;
	Chen,~X.; Yao,~W.; Cui,~X. \emph{Scientific Reports} \textbf{2013},
	\emph{3}\relax
	\mciteBstWouldAddEndPuncttrue
	\mciteSetBstMidEndSepPunct{\mcitedefaultmidpunct}
	{\mcitedefaultendpunct}{\mcitedefaultseppunct}\relax
	\EndOfBibitem
	\bibitem[Xu \latin{et~al.}(2014)Xu, Yao, Xiao, and Heinz]{xu_spin_2014}
	Xu,~X.; Yao,~W.; Xiao,~D.; Heinz,~T.~F. \emph{Nature Physics} \textbf{2014},
	\emph{10}, 343--350\relax
	\mciteBstWouldAddEndPuncttrue
	\mciteSetBstMidEndSepPunct{\mcitedefaultmidpunct}
	{\mcitedefaultendpunct}{\mcitedefaultseppunct}\relax
	\EndOfBibitem
	\bibitem[Mak \latin{et~al.}(2014)Mak, McGill, Park, and
	McEuen]{mak_valley_2014}
	Mak,~K.~F.; McGill,~K.~L.; Park,~J.; McEuen,~P.~L. \emph{Science}
	\textbf{2014}, \emph{344}, 1489--1492\relax
	\mciteBstWouldAddEndPuncttrue
	\mciteSetBstMidEndSepPunct{\mcitedefaultmidpunct}
	{\mcitedefaultendpunct}{\mcitedefaultseppunct}\relax
	\EndOfBibitem
	\bibitem[Zhu \latin{et~al.}(2011)Zhu, Cheng, and
	Schwingenschlögl]{zhu_giant_2011}
	Zhu,~Z.~Y.; Cheng,~Y.~C.; Schwingenschlögl,~U. \emph{Physical Review B}
	\textbf{2011}, \emph{84}, 153402\relax
	\mciteBstWouldAddEndPuncttrue
	\mciteSetBstMidEndSepPunct{\mcitedefaultmidpunct}
	{\mcitedefaultendpunct}{\mcitedefaultseppunct}\relax
	\EndOfBibitem
	\bibitem[Ye \latin{et~al.}(2014)Ye, Cao, O’Brien, Zhu, Yin, Wang, Louie, and
	Zhang]{ye_probing_2014}
	Ye,~Z.; Cao,~T.; O’Brien,~K.; Zhu,~H.; Yin,~X.; Wang,~Y.; Louie,~S.~G.;
	Zhang,~X. \emph{Nature} \textbf{2014}, \emph{513}, 214--218\relax
	\mciteBstWouldAddEndPuncttrue
	\mciteSetBstMidEndSepPunct{\mcitedefaultmidpunct}
	{\mcitedefaultendpunct}{\mcitedefaultseppunct}\relax
	\EndOfBibitem
	\bibitem[Chernikov \latin{et~al.}(2014)Chernikov, Berkelbach, Hill, Rigosi, Li,
	Aslan, Reichman, Hybertsen, and Heinz]{chernikov_exciton_2014}
	Chernikov,~A.; Berkelbach,~T.~C.; Hill,~H.~M.; Rigosi,~A.; Li,~Y.;
	Aslan,~O.~B.; Reichman,~D.~R.; Hybertsen,~M.~S.; Heinz,~T.~F. \emph{Physical
		Review Letters} \textbf{2014}, \emph{113}, 076802\relax
	\mciteBstWouldAddEndPuncttrue
	\mciteSetBstMidEndSepPunct{\mcitedefaultmidpunct}
	{\mcitedefaultendpunct}{\mcitedefaultseppunct}\relax
	\EndOfBibitem
	\bibitem[Kozawa \latin{et~al.}(2014)Kozawa, Kumar, Carvalho, Kumar~Amara, Zhao,
	Wang, Toh, Ribeiro, Castro~Neto, Matsuda, and Eda]{kozawa_photocarrier_2014}
	Kozawa,~D.; Kumar,~R.; Carvalho,~A.; Kumar~Amara,~K.; Zhao,~W.; Wang,~S.;
	Toh,~M.; Ribeiro,~R.~M.; Castro~Neto,~A.~H.; Matsuda,~K.; Eda,~G.
	\emph{Nature Communications} \textbf{2014}, \emph{5}\relax
	\mciteBstWouldAddEndPuncttrue
	\mciteSetBstMidEndSepPunct{\mcitedefaultmidpunct}
	{\mcitedefaultendpunct}{\mcitedefaultseppunct}\relax
	\EndOfBibitem
	\bibitem[Chernikov \latin{et~al.}(2015)Chernikov, Ruppert, Hill, Rigosi, and
	Heinz]{chernikov_population_2015}
	Chernikov,~A.; Ruppert,~C.; Hill,~H.~M.; Rigosi,~A.~F.; Heinz,~T.~F.
	\emph{Nature Photonics} \textbf{2015}, \emph{9}, 466--470\relax
	\mciteBstWouldAddEndPuncttrue
	\mciteSetBstMidEndSepPunct{\mcitedefaultmidpunct}
	{\mcitedefaultendpunct}{\mcitedefaultseppunct}\relax
	\EndOfBibitem
	\bibitem[Yu \latin{et~al.}(2014)Yu, Liu, Gong, Xu, and Yao]{yu_dirac_2014}
	Yu,~H.; Liu,~G.-B.; Gong,~P.; Xu,~X.; Yao,~W. \emph{Nature Communications}
	\textbf{2014}, \emph{5}, 3876\relax
	\mciteBstWouldAddEndPuncttrue
	\mciteSetBstMidEndSepPunct{\mcitedefaultmidpunct}
	{\mcitedefaultendpunct}{\mcitedefaultseppunct}\relax
	\EndOfBibitem
	\bibitem[Mak \latin{et~al.}(2013)Mak, He, Lee, Lee, Hone, Heinz, and
	Shan]{mak_tightly_2013}
	Mak,~K.~F.; He,~K.; Lee,~C.; Lee,~G.~H.; Hone,~J.; Heinz,~T.~F.; Shan,~J.
	\emph{Nature Materials} \textbf{2013}, \emph{12}, 207--211\relax
	\mciteBstWouldAddEndPuncttrue
	\mciteSetBstMidEndSepPunct{\mcitedefaultmidpunct}
	{\mcitedefaultendpunct}{\mcitedefaultseppunct}\relax
	\EndOfBibitem
	\bibitem[Jones \latin{et~al.}(2013)Jones, Yu, Ghimire, Wu, Aivazian, Ross,
	Zhao, Yan, Mandrus, Xiao, Yao, and Xu]{jones_optical_2013}
	Jones,~A.~M.; Yu,~H.; Ghimire,~N.~J.; Wu,~S.; Aivazian,~G.; Ross,~J.~S.;
	Zhao,~B.; Yan,~J.; Mandrus,~D.~G.; Xiao,~D.; Yao,~W.; Xu,~X. \emph{Nature
		Nanotechnology} \textbf{2013}, \emph{8}, 634--638\relax
	\mciteBstWouldAddEndPuncttrue
	\mciteSetBstMidEndSepPunct{\mcitedefaultmidpunct}
	{\mcitedefaultendpunct}{\mcitedefaultseppunct}\relax
	\EndOfBibitem
	\bibitem[Mak \latin{et~al.}(2012)Mak, He, Shan, and Heinz]{mak_control_2012}
	Mak,~K.~F.; He,~K.; Shan,~J.; Heinz,~T.~F. \emph{Nature Nanotechnology}
	\textbf{2012}, \emph{7}, 494--498\relax
	\mciteBstWouldAddEndPuncttrue
	\mciteSetBstMidEndSepPunct{\mcitedefaultmidpunct}
	{\mcitedefaultendpunct}{\mcitedefaultseppunct}\relax
	\EndOfBibitem
	\bibitem[Zeng \latin{et~al.}(2012)Zeng, Dai, Yao, Xiao, and
	Cui]{zeng_valley_2012}
	Zeng,~H.; Dai,~J.; Yao,~W.; Xiao,~D.; Cui,~X. \emph{Nature Nanotechnology}
	\textbf{2012}, \emph{7}, 490--493\relax
	\mciteBstWouldAddEndPuncttrue
	\mciteSetBstMidEndSepPunct{\mcitedefaultmidpunct}
	{\mcitedefaultendpunct}{\mcitedefaultseppunct}\relax
	\EndOfBibitem
	\bibitem[Cao \latin{et~al.}(2012)Cao, Wang, Han, Ye, Zhu, Shi, Niu, Tan, Wang,
	Liu, and Feng]{cao_valley-selective_2012}
	Cao,~T.; Wang,~G.; Han,~W.; Ye,~H.; Zhu,~C.; Shi,~J.; Niu,~Q.; Tan,~P.;
	Wang,~E.; Liu,~B.; Feng,~J. \emph{Nature Communications} \textbf{2012},
	\emph{3}, 887\relax
	\mciteBstWouldAddEndPuncttrue
	\mciteSetBstMidEndSepPunct{\mcitedefaultmidpunct}
	{\mcitedefaultendpunct}{\mcitedefaultseppunct}\relax
	\EndOfBibitem
	\bibitem[Yu \latin{et~al.}(2015)Yu, Wang, Tong, Xu, and Yao]{yu_anomalous_2015}
	Yu,~H.; Wang,~Y.; Tong,~Q.; Xu,~X.; Yao,~W. \emph{Physical Review Letters}
	\textbf{2015}, \emph{115}, 187002\relax
	\mciteBstWouldAddEndPuncttrue
	\mciteSetBstMidEndSepPunct{\mcitedefaultmidpunct}
	{\mcitedefaultendpunct}{\mcitedefaultseppunct}\relax
	\EndOfBibitem
	\bibitem[Wu \latin{et~al.}(2013)Wu, Jariwala, Sangwan, Marks, Hersam, and
	Lauhon]{wu_elucidating_2013}
	Wu,~C.-C.; Jariwala,~D.; Sangwan,~V.~K.; Marks,~T.~J.; Hersam,~M.~C.;
	Lauhon,~L.~J. \emph{The Journal of Physical Chemistry Letters} \textbf{2013},
	\emph{4}, 2508--2513\relax
	\mciteBstWouldAddEndPuncttrue
	\mciteSetBstMidEndSepPunct{\mcitedefaultmidpunct}
	{\mcitedefaultendpunct}{\mcitedefaultseppunct}\relax
	\EndOfBibitem
	\bibitem[Ubrig \latin{et~al.}(2014)Ubrig, Jo, Berger, Morpurgo, and
	Kuzmenko]{ubrig_scanning_2014}
	Ubrig,~N.; Jo,~S.; Berger,~H.; Morpurgo,~A.~F.; Kuzmenko,~A.~B. \emph{Applied
		Physics Letters} \textbf{2014}, \emph{104}, 171112\relax
	\mciteBstWouldAddEndPuncttrue
	\mciteSetBstMidEndSepPunct{\mcitedefaultmidpunct}
	{\mcitedefaultendpunct}{\mcitedefaultseppunct}\relax
	\EndOfBibitem
	\bibitem[Ahn \latin{et~al.}(2005)Ahn, Dunning, and Park]{ahn_scanning_2005}
	Ahn,~Y.; Dunning,~J.; Park,~J. \emph{Nano Letters} \textbf{2005}, \emph{5},
	1367--1370\relax
	\mciteBstWouldAddEndPuncttrue
	\mciteSetBstMidEndSepPunct{\mcitedefaultmidpunct}
	{\mcitedefaultendpunct}{\mcitedefaultseppunct}\relax
	\EndOfBibitem
	\bibitem[Yamaguchi \latin{et~al.}(2015)Yamaguchi, Blancon, Kappera, Lei,
	Najmaei, Mangum, Gupta, Ajayan, Lou, Chhowalla, Crochet, and
	Mohite]{yamaguchi_spatially_2015}
	Yamaguchi,~H.; Blancon,~J.-C.; Kappera,~R.; Lei,~S.; Najmaei,~S.;
	Mangum,~B.~D.; Gupta,~G.; Ajayan,~P.~M.; Lou,~J.; Chhowalla,~M.;
	Crochet,~J.~J.; Mohite,~A.~D. \emph{ACS Nano} \textbf{2015}, \emph{9},
	840--849\relax
	\mciteBstWouldAddEndPuncttrue
	\mciteSetBstMidEndSepPunct{\mcitedefaultmidpunct}
	{\mcitedefaultendpunct}{\mcitedefaultseppunct}\relax
	\EndOfBibitem
	\bibitem[Wang \latin{et~al.}(2012)Wang, Ruzicka, Kumar, Bellus, Chiu, and
	Zhao]{wang_ultrafast_2012}
	Wang,~R.; Ruzicka,~B.~A.; Kumar,~N.; Bellus,~M.~Z.; Chiu,~H.-Y.; Zhao,~H.
	\emph{Physical Review B} \textbf{2012}, \emph{86}, 045406\relax
	\mciteBstWouldAddEndPuncttrue
	\mciteSetBstMidEndSepPunct{\mcitedefaultmidpunct}
	{\mcitedefaultendpunct}{\mcitedefaultseppunct}\relax
	\EndOfBibitem
	\bibitem[Moody \latin{et~al.}(2016)Moody, Schaibley, and
	Xu]{moody_exciton_2016}
	Moody,~G.; Schaibley,~J.; Xu,~X. \emph{JOSA B} \textbf{2016}, \emph{33},
	C39--C49\relax
	\mciteBstWouldAddEndPuncttrue
	\mciteSetBstMidEndSepPunct{\mcitedefaultmidpunct}
	{\mcitedefaultendpunct}{\mcitedefaultseppunct}\relax
	\EndOfBibitem
	\bibitem[Jo \latin{et~al.}(2015)Jo, Costanzo, Berger, and
	Morpurgo]{jo_electrostatically_2015}
	Jo,~S.; Costanzo,~D.; Berger,~H.; Morpurgo,~A.~F. \emph{Nano Letters}
	\textbf{2015}, \emph{15}, 1197--1202\relax
	\mciteBstWouldAddEndPuncttrue
	\mciteSetBstMidEndSepPunct{\mcitedefaultmidpunct}
	{\mcitedefaultendpunct}{\mcitedefaultseppunct}\relax
	\EndOfBibitem
	\bibitem[Costanzo \latin{et~al.}(2016)Costanzo, Jo, Berger, and
	Morpurgo]{costanzo_gate-induced_2016}
	Costanzo,~D.; Jo,~S.; Berger,~H.; Morpurgo,~A.~F. \emph{Nature Nanotechnology}
	\textbf{2016}, \emph{11}, 339--344\relax
	\mciteBstWouldAddEndPuncttrue
	\mciteSetBstMidEndSepPunct{\mcitedefaultmidpunct}
	{\mcitedefaultendpunct}{\mcitedefaultseppunct}\relax
	\EndOfBibitem
	\bibitem[Braga \latin{et~al.}(2012)Braga, Gutiérrez~Lezama, Berger, and
	Morpurgo]{braga_quantitative_2012}
	Braga,~D.; Gutiérrez~Lezama,~I.; Berger,~H.; Morpurgo,~A.~F. \emph{Nano
		Letters} \textbf{2012}, \emph{12}, 5218--5223\relax
	\mciteBstWouldAddEndPuncttrue
	\mciteSetBstMidEndSepPunct{\mcitedefaultmidpunct}
	{\mcitedefaultendpunct}{\mcitedefaultseppunct}\relax
	\EndOfBibitem
	\bibitem[Ponomarev \latin{et~al.}(2015)Ponomarev, Gutiérrez-Lezama, Ubrig, and
	Morpurgo]{ponomarev_ambipolar_2015}
	Ponomarev,~E.; Gutiérrez-Lezama,~I.; Ubrig,~N.; Morpurgo,~A.~F. \emph{Nano
		Letters} \textbf{2015}, \emph{15}, 8289--8294\relax
	\mciteBstWouldAddEndPuncttrue
	\mciteSetBstMidEndSepPunct{\mcitedefaultmidpunct}
	{\mcitedefaultendpunct}{\mcitedefaultseppunct}\relax
	\EndOfBibitem
	\bibitem[Jo \latin{et~al.}(2014)Jo, Ubrig, Berger, Kuzmenko, and
	Morpurgo]{jo_mono-_2014}
	Jo,~S.; Ubrig,~N.; Berger,~H.; Kuzmenko,~A.~B.; Morpurgo,~A.~F. \emph{Nano
		Letters} \textbf{2014}, \emph{14}, 2019--2025\relax
	\mciteBstWouldAddEndPuncttrue
	\mciteSetBstMidEndSepPunct{\mcitedefaultmidpunct}
	{\mcitedefaultendpunct}{\mcitedefaultseppunct}\relax
	\EndOfBibitem
	\bibitem[Not()]{Note-1}
	In simple terms, Hall probes are metal-semiconductor Schottky contacts behaving
	as diodes, and they can in principle produce a dc signal by rectifying the
	oscillating high-frequency currents produced by the incident light. An
	asymmetry in the contact would then lead to a voltage difference between the
	Hall probes that could be mistakenly interpreted as a manifestation of the
	VHE.\relax
	\mciteBstWouldAddEndPunctfalse
	\mciteSetBstMidEndSepPunct{\mcitedefaultmidpunct}
	{}{\mcitedefaultseppunct}\relax
	\EndOfBibitem
	\bibitem[Mitioglu \latin{et~al.}(2013)Mitioglu, Plochocka, Jadczak, Escoffier,
	Rikken, Kulyuk, and Maude]{mitioglu_optical_2013}
	Mitioglu,~A.~A.; Plochocka,~P.; Jadczak,~J.~N.; Escoffier,~W.; Rikken,~G. L.
	J.~A.; Kulyuk,~L.; Maude,~D.~K. \emph{Physical Review B} \textbf{2013},
	\emph{88}, 245403\relax
	\mciteBstWouldAddEndPuncttrue
	\mciteSetBstMidEndSepPunct{\mcitedefaultmidpunct}
	{\mcitedefaultendpunct}{\mcitedefaultseppunct}\relax
	\EndOfBibitem
	\bibitem[Cadiz \latin{et~al.}(2017)Cadiz, Courtade, Robert, Wang, Shen, Cai,
	Taniguchi, Watanabe, Carrere, Lagarde, Manca, Amand, Renucci, Tongay, Marie,
	and Urbaszek]{cadiz_excitonic_2017}
	Cadiz,~F. \latin{et~al.}  \emph{Physical Review X} \textbf{2017}, \emph{7},
	021026\relax
	\mciteBstWouldAddEndPuncttrue
	\mciteSetBstMidEndSepPunct{\mcitedefaultmidpunct}
	{\mcitedefaultendpunct}{\mcitedefaultseppunct}\relax
	\EndOfBibitem
	\bibitem[Pei \latin{et~al.}(2015)Pei, Yang, Xu, Zeng, Myint, Zhang, Zheng, Qin,
	Wang, Jiang, and Lu]{pei_exciton_2015}
	Pei,~J.; Yang,~J.; Xu,~R.; Zeng,~Y.-H.; Myint,~Y.~W.; Zhang,~S.; Zheng,~J.-C.;
	Qin,~Q.; Wang,~X.; Jiang,~W.; Lu,~Y. \emph{Small} \textbf{2015}, \emph{11},
	6384--6390\relax
	\mciteBstWouldAddEndPuncttrue
	\mciteSetBstMidEndSepPunct{\mcitedefaultmidpunct}
	{\mcitedefaultendpunct}{\mcitedefaultseppunct}\relax
	\EndOfBibitem
	\bibitem[Suzuki \latin{et~al.}(2014)Suzuki, Sakano, Zhang, Akashi, Morikawa,
	Harasawa, Yaji, Kuroda, Miyamoto, Okuda, Ishizaka, Arita, and
	Iwasa]{suzuki_valley-dependent_2014}
	Suzuki,~R.; Sakano,~M.; Zhang,~Y.~J.; Akashi,~R.; Morikawa,~D.; Harasawa,~A.;
	Yaji,~K.; Kuroda,~K.; Miyamoto,~K.; Okuda,~T.; Ishizaka,~K.; Arita,~R.;
	Iwasa,~Y. \emph{Nature Nanotechnology} \textbf{2014}, \emph{9},
	611--617\relax
	\mciteBstWouldAddEndPuncttrue
	\mciteSetBstMidEndSepPunct{\mcitedefaultmidpunct}
	{\mcitedefaultendpunct}{\mcitedefaultseppunct}\relax
	\EndOfBibitem
	\bibitem[Lee \latin{et~al.}(2016)Lee, Mak, and Shan]{lee_electrical_2016}
	Lee,~J.; Mak,~K.~F.; Shan,~J. \emph{Nature Nanotechnology} \textbf{2016},
	\emph{11}, 421--425\relax
	\mciteBstWouldAddEndPuncttrue
	\mciteSetBstMidEndSepPunct{\mcitedefaultmidpunct}
	{\mcitedefaultendpunct}{\mcitedefaultseppunct}\relax
	\EndOfBibitem
	\bibitem[Wu \latin{et~al.}(2013)Wu, Ross, Liu, Aivazian, Jones, Fei, Zhu, Xiao,
	Yao, Cobden, and Xu]{wu_electrical_2013}
	Wu,~S.; Ross,~J.~S.; Liu,~G.-B.; Aivazian,~G.; Jones,~A.; Fei,~Z.; Zhu,~W.;
	Xiao,~D.; Yao,~W.; Cobden,~D.; Xu,~X. \emph{Nature Physics} \textbf{2013},
	\emph{9}, 149--153\relax
	\mciteBstWouldAddEndPuncttrue
	\mciteSetBstMidEndSepPunct{\mcitedefaultmidpunct}
	{\mcitedefaultendpunct}{\mcitedefaultseppunct}\relax
	\EndOfBibitem
	\bibitem[Beal \latin{et~al.}(1972)Beal, Knights, and
	Liang]{beal_transmission_1972}
	Beal,~A.~R.; Knights,~J.~C.; Liang,~W.~Y. \emph{Journal of Physics C: Solid
		State Physics} \textbf{1972}, \emph{5}, 3540\relax
	\mciteBstWouldAddEndPuncttrue
	\mciteSetBstMidEndSepPunct{\mcitedefaultmidpunct}
	{\mcitedefaultendpunct}{\mcitedefaultseppunct}\relax
	\EndOfBibitem
	\bibitem[Beal \latin{et~al.}(1976)Beal, Liang, and
	Hughes]{beal_kramers-kronig_1976}
	Beal,~A.~R.; Liang,~W.~Y.; Hughes,~H.~P. \emph{Journal of Physics C: Solid
		State Physics} \textbf{1976}, \emph{9}, 2449--2457\relax
	\mciteBstWouldAddEndPuncttrue
	\mciteSetBstMidEndSepPunct{\mcitedefaultmidpunct}
	{\mcitedefaultendpunct}{\mcitedefaultseppunct}\relax
	\EndOfBibitem
	\bibitem[Zhao \latin{et~al.}(2013)Zhao, Ghorannevis, Chu, Toh, Kloc, Tan, and
	Eda]{zhao_evolution_2013}
	Zhao,~W.; Ghorannevis,~Z.; Chu,~L.; Toh,~M.; Kloc,~C.; Tan,~P.-H.; Eda,~G.
	\emph{ACS Nano} \textbf{2013}, \emph{7}, 791--797\relax
	\mciteBstWouldAddEndPuncttrue
	\mciteSetBstMidEndSepPunct{\mcitedefaultmidpunct}
	{\mcitedefaultendpunct}{\mcitedefaultseppunct}\relax
	\EndOfBibitem
\end{mcitethebibliography}
\pagebreak


\end{document}